\documentclass[onecolumn,superscriptaddress,aps]{revtex4}

\usepackage{amsmath}
\usepackage{graphicx}

\begin{document}

\title{Canonical active Brownian motion}

\author{Alexander Gl\"uck} \email{alexander.glueck@chello.at}
\affiliation{Faculty of Physics, University of Vienna,
   Boltzmanngasse 5, A-1090 Wien}
\author{Helmuth H\"uffel} 
\affiliation{Faculty of Physics, University of Vienna,
   Boltzmanngasse 5, A-1090 Wien}
\author{Sa{\v s}a Iliji{\'c}} 
\affiliation{Department of Physics,
   Faculty of Electrical Engineering and Computing,
   University of Zagreb, Unska 3, HR-10000 Zagreb}

\begin{abstract}
Active Brownian motion is the complex motion of active Brownian particles.
They are ``active'' in the sense that they can transform
their internal energy into energy of motion
and thus create complex motion patterns.
Theories of active Brownian motion so far imposed couplings
between the internal energy and the \emph{kinetic} energy of the system.
We investigate how this idea can be naturally taken further
to include also couplings to the \emph{potential} energy,
which finally leads to a general theory of canonical dissipative systems.
Explicit analytical and numerical studies are done
for the motion of one particle in harmonic external potentials.
Apart from stationary solutions, we study non-equilibrium dynamics
and show the existence of various bifurcation phenomena. 
\end{abstract}

\maketitle

\section{Preliminaries}

The theory of Brownian motion in its formulation due to Langevin \cite{key-5}
assumes that particles are subject to stochastic influences
and external forces, the latter making them move
according to the potential $U(q)$,
which models their environmental landscape.
Stochastic equations of motion are
  \begin{equation}
  \frac{dq_{i}}{dt} = p_{i}\quad,\quad\frac{dp_{i}}{dt}
  = - \gamma_{0}p_{i}-\frac{\partial U}{\partial q_{i}}+\eta_{i},
  \end{equation}
including a friction term with friction constant $\gamma_{0}>0$,
the force of the external potential $-\partial_{i}U$
and the random force $\eta_{i}$,
which has infinite variance per definition.
Particles are considered to be without an inner structure,
just driven by the external potential and the noise,
without any capability of changing their dynamics by themselves.
If studying the complex motion e.g.\ of bacteria
or even higher developed organisms,
this simple assumption of particles
reacting to a prescribed potential landscape
cannot explain the widely observed emergent phenomena
arising in such systems.
A number of models were proposed,
including some kind of `self driven motion' of particles,
which should account for the lack of complex behavior dynamics.
Apart from just postulating such additional effects,
the theory of active Brownian particles \cite{key-21,key-1,key-2,buch}
explains the origin of this `self-motion' by imposing an additional
\emph{internal} degree of freedom, called `internal energy' $e$.
This energy can be increased by taking up external energy (`food')
from the environment and be transformed into kinetic energy of the particle,
i.e. the particle is `active' in the sense
that it is able to convert its internal energy into energy of motion.
This model has served to describe animal mobility in general \cite{mobi}
and was able to make quantitative statements
about emergent self-organized properties
in the collective behavior of many particle systems,
e.g.\ swarms \cite{mycitation,key-3,key-4}.

After reviewing some fundamental aspects of the original theory,
we propose a canonical version of active Brownian motion
by allowing to convert internal energy
into the full mechanical energy of the particle.
In case of stationary internal energies
our generalized theory defines active Brownian motion
as a canonical dissipative system also with interactions,
which was not possible in former models.
Interacting active Brownian particles were studied intensively
\cite{key-21,key-1,key-2,buch,mobi,mycitation},
recent contributions included dissipative Toda-- and Morse--systems
\cite{toda,Eb18,Ch19}.
In our formalism, however, we are able to study interacting active
Brownian motion as a \emph{canonical} dissipative system.

The original formulations of active Brownian motion
rarely discuss nonequilibrium dynamics.
In most applications, only stationary solutions are present,
which implies constant internal energies for all times.
We will show how active Brownian motion in its fully coupled form
contains rich nonequilibrium structures, too.

\section{Original theory:
Coupling of $e$ to the kinetic energy \label{sec:orig}}

The main idea of active Brownian motion is to ascribe an additional,
so called `internal' energy $e$ \cite{buch} to Brownian particles.
Stochastic dynamics for the motion of an active particle in $d$ dimensions
with degrees of freedom $q_{i}$ and associated momenta $p_{i}$,
where $i=1,\ldots,d$ and $\mathbf{q}=(q_{i})$, $\mathbf{p}=(p_{i})$
read \cite{buch}:
  \begin{equation}
  \frac{dq_{i}}{dt} = \frac{\partial H}{\partial p_{i}}
  \end{equation}
  \begin{equation}
  \frac{dp_{i}}{dt} = -\frac{\partial H}{\partial q_{i}}-g(e)
  \frac{\partial H}{\partial p_{i}}+\eta_{i} \label{eq:kafka}
  \end{equation}
  \begin{equation}
  \frac{de}{dt} = c_{1}-c_{2}e-c_{3}e\frac{p^{2}}{2} \label{hui}
  \end{equation}
with $H(q,p)=\frac{p^{2}}{2}+U(q)$ and $g(e)=\gamma_{0}-d_{2}e$.
Noise correlations are
  \begin{equation}
  \langle\eta_{i}(t)\rangle
  =0\quad,\quad\langle\eta_{i}(t)\eta_{j}(\bar{t})\rangle
  =2\delta_{ij}\delta(t-\bar{t}).
  \end{equation}
All constants $c_{i}$, as well as $\gamma_{0},\, d_{2}$
are assumed to be positive.
Calculating the total time derivative of the Hamiltonian
in the deterministic case yields
  \begin{equation}
  \frac{dH}{dt}=-g(e)\left(\frac{\partial H}{\partial\mathbf{p}}\right)^{2}
  =-\gamma_{0}p^{2}+d_{2}ep^{2},
  \end{equation}
which allows us to interpret the implications of internal energy dynamics:
The first term models the mechanical energy loss by friction
due to the surrounding media,
whereas the second term shows the possibility of an energy increase
via the coupling of kinetic energy to the internal energy.
If the internal energy becomes stationary after some time ($\frac{de}{dt}=0$),
it can be expressed as a function of the kinetic energy:
  \begin{equation}
  e=\frac{c_{1}}{c_{2}+c_{3}\frac{p^{2}}{2}},
  \end{equation}
which implies
  \begin{equation}
  \frac{dp_{i}}{dt}
  =-\frac{\partial H}{\partial q_{i}}-\gamma(p)
      \frac{\partial H}{\partial p_{i}}+\eta_{i},
  \label{anders}
  \end{equation}
where $\gamma(p)=\gamma_{0}-\frac{d_{2}c_{1}}{c_{2}+c_{3}\frac{p^{2}}{2}}$.
The equation above defines a model of Brownian motion
in media of nonlinear friction.
Active Brownian particles with stationary internal energies
therefore move like in a medium with nonlinear friction function $\gamma(p)$.
Relaxation of the internal energy to a stationary value
therefore results in changing the original friction of the medium
to an effective friction.
The following section will show that equilibrium internal energies
will have completely different effects in the case
when $e$ is coupled to the potential energy $U(q)$,
instead of coupling it to $\frac{p^{2}}{2}$, as is usually done.
Setting $\frac{de}{dt}=0$ will then result
in changing the original potential to an\emph{ }effective one.

\section{Coupling of $e$ to the potential energy \label{sec:potential coup}}

Active Brownian motion in its original formulation
imposes a coupling of the internal to the kinetic energy,
given by equation (\ref{hui}),
in which $e$ is multiplied by $\frac{p^{2}}{2}$.
The most natural and simplest way to extend this balance equation
to the case of potential couplings would be to multiply $e$ with $U(q)$.
Furthermore, in equation (\ref{eq:kafka}),
it should appear the product between the internal energy
and the $\mathbf{q}$-gradient of $H$
(and not the $\mathbf{p}$-gradient as in the original case).
Especially when considering many particle systems like swarms,
exchanges between internal and potential energies seem physically intriguing:
Swarm particles generate their interaction potentials mutually
due to exchange of internal energy;
the internal energy of one particle
has effects on its interaction with all the other particles.
In the present paper we study the dynamics of a single particle only,
but we already introduce its generalized couplings for the internal energy.

We have already given stochastic differential equations
for this set-up in configuration space \cite{key-6}.
Stochastic dynamics in phase space equivalently read:
  \begin{equation}
  \frac{dq_{i}}{dt} = \frac{\partial H}{\partial p_{i}} \label{eq:lop}
  \end{equation}
  \begin{equation}
  \frac{dp_{i}}{dt} = -f(e)\frac{\partial H}{\partial q_{i}}-\gamma_{0}
      \frac{\partial H}{\partial p_{i}}+\eta_{i}
  \label{beckett}
  \end{equation}
  \begin{equation}
  \frac{de}{dt} = c_{1}-c_{2}e-c_{3}eU\label{sop}
  \end{equation}
with $H(q,p)=\frac{p^{2}}{2}+U(q)$ and $f(e)=1-d_{1}e$.
The system is constructed in complete analogy
to the original theory of active Brownian motion,
but with interchanging the role of $\mathbf{p}$ and $\mathbf{q}$,
resp.\ $\frac{p^{2}}{2}$ and $U(q)$.
The evolution equation for the Hamiltonian
is calculated in the same way as in Sec.\ \ref{sec:orig}, which leads to
  \begin{equation}
  \dot H = - \gamma_0 \, p^2 + d_1 e \, (\mathbf{p} \cdot
  \mbox{\boldmath$\nabla$} U).
  \end{equation}
Discussions about the interpretation of such conservation laws
can be found in the standard textbook \cite{buch}.

\begin{figure}[t]
\begin{center}
\includegraphics{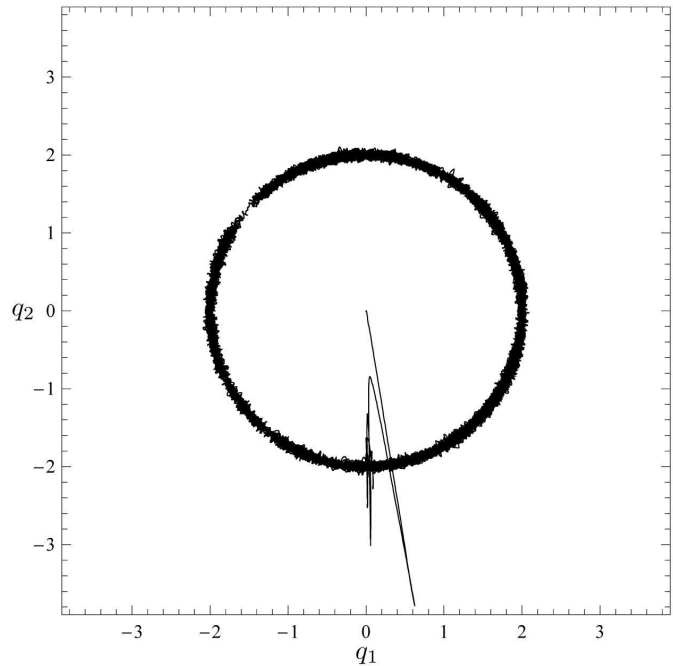}
\end{center}
\caption{\label{fig:diffusion circle}
Two dimensional motion of an active particle
under the influence of harmonic forces in the $q_{1,}q_{2}$-plane,
subject to equations (\ref{eq:lop})-(\ref{sop})
with all parameters chosen to be equal to one
except $d_{1}=5$ and $s^{\frac{1}{2}}=0.3$.
The radius of the circle can be calculated
from equation (\ref{eq:radius}): $r_{0}=2$.
Initial conditions are
$q_{1}(0)=0=q_{2}(0),\, p_{1}(0)=0=p_{2}(0),\, e(0)=0$.}
\end{figure}

One remarkable feature is the emergence of effective potentials
in the time region of stationary internal energy, i.e. $\frac{de}{dt}=0$;
hence
  \begin{equation}
  e(q)=\frac{c_{1}}{c_{2}+c_{3}U(q)}
  \label{stationary}
  \end{equation}
and thus
  \begin{equation}
  \frac{dq_{i}}{dt} = \frac{\partial\widetilde{H}}{\partial p_{i}} \label{loee}
  \end{equation}
  \begin{equation}
  \frac{dp_{i}}{dt} = -\frac{\partial\tilde{H}}{\partial q_{i}}
    -\gamma_{0}\frac{\partial\tilde{H}}{\partial p_{i}}+\eta_{i}, \label{eloo}
  \end{equation}
which are formally the well known equations
for the motion of a Brownian particle in media of linear friction,
but with an effective Hamiltonian
  \begin{equation}
  \tilde{H}=\frac{p^{2}}{2}+\tilde{U}(q)
  \quad,\quad
  \tilde{U}(q)=U(q)-\frac{d_{1}c_{1}}{c_{3}}\ln\left(c_{2}+c_{3}U(q)\right).
  \label{eq:effective}
  \end{equation}
Its time evolution in the deterministic case is given by
  \begin{equation}
  \frac{d\tilde{H}}{dt}=-\gamma_{0}p^{2}.
  \end{equation}
Couplings of $e$ to the potential energy therefore manifest themselves
by changing the original potential to an effective one.
Stationary internal energies in classical active Brownian motion
describe systems of a particle moving
with an effective nonlinear friction $\gamma(p)$,
whereas in the case of internal energy coupled to $U(q)$,
the arising effect lies in changing the original potential
to an effective one $\tilde{U}(q)$.
The crucial role of this effective potential
can be seen by numerical investigations of equations (\ref{eq:lop})-(\ref{sop})
for specific forms of $U(q)$:
Under the influence of harmonic forces $U=aq^{2}$,
the motion of one particle in two dimensions is given by
  \begin{equation}
  \frac{dq_{1}}{dt} = p_{1}\quad,\quad\frac{dq_{2}}{dt}=p_{2}
  \label{wald}
  \end{equation}
  \begin{equation}
  \frac{dp_{1}}{dt} = -\gamma_{0\,}p_{1}-2a\,
    q_{1}+2d_{1}\, e\, a\, q_{1}+s^{\frac{1}{2}}\,\xi_{1}
  \label{waldig}
  \end{equation}
  \begin{equation}
  \frac{dp_{2}}{dt} = -\gamma_{0\,}p_{2}-2a\, q_{2}+2d_{1\,}e\, a\,
    q_{2}+s^{\frac{1}{2}}\,\xi_{2}
  \label{waldiger}
  \end{equation}
  \begin{equation}
  \frac{de}{dt} = c_{1}-c_{2}\, e-c_{3}\, e\, a
    \left(q_{1}{}^{2}+q_{2}{}^{2}\right)
  \label{waldvoll}
  \end{equation}
where a noise strength factor $s$
was introduced via the substitution $\eta=s^{\frac{1}{2}}\xi$.
(For a detailed discussion of equilibrium solutions
and its bifurcations, see Section 5.)
Figure \ref{fig:diffusion circle} shows the simulated behavior
of a particle subject to the equations given above.

\begin{figure}
\begin{center}
\includegraphics{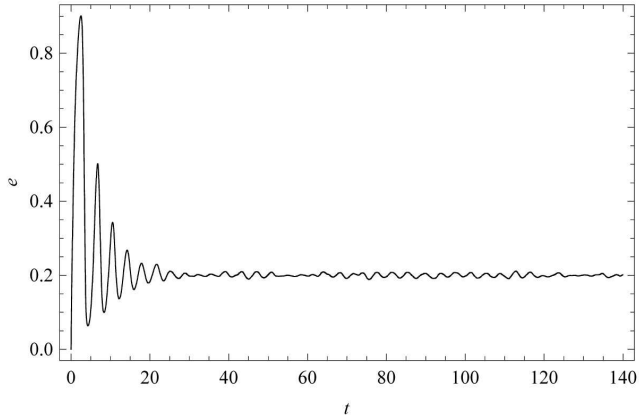}
\end{center}
\caption{\label{fig:Internal-energy-evolution}
Internal energy evolution in time corresponding to the parameter setup
given in Figure \ref{fig:diffusion circle}.
The stationary value of $e$ can be calculated
from equation (\ref{e0}): $e_{0}=1/5$.}
\end{figure}

Out of its initial conditions,
the particle drops onto a circle and then diffuses in no preferred direction,
until the whole ring area is filled with trajectories.
This random motion on a ring can be explained
by referring to the structure of the effective potential,
which for the harmonic example reads
  \begin{equation}
  \tilde{U}(r)=ar^{2}-\frac{d_{1}c_{1}}{c_{3}}
      \ln\left(c_{2}+c_{3}ar^{2}\right)
  \end{equation}
in terms of the radial coordinate
$r=\left(q_{1}{}^{2}+q_{2}{}^{2}\right)^{\frac{1}{2}}$.
The minima of $\tilde{U}(r)$ are found to be at all radial distances
  \begin{equation}
  r_{0}=\left(\frac{d_{1}c_{1}-c_{2}}{ac_{3}}\right)^{\frac{1}{2}},
  \label{eq:radius}
  \end{equation}
which is exactly the radius of the circle
seen in Figure \ref{fig:diffusion circle}.
In the case of no coupling to an internal energy ($d_{1}=0$),
the particle would directly fall into the minimum of the external potential
(which is at $r=0$), and no diffusion in any direction would be present.
Positive values of $r_{0}$ can be realized
for critical values of the coupling parameter $d_{1}$,
namely if $d_{1}>\frac{c_{2}}{c_{1}}$.
Active Brownian particles with a coupling of $e$
to the potential energy feel a different landscape than the prescribed one,
i.e.\ an \emph{effective} environment,
which is then searched for minimum areas.
The internal energy effect in our present example
results in trapping the particle not at the origin,
but on a circle of a given radius $r=r_{0}$.
This stationary state of noise-induced wandering
in the valley of minima is immediately present
after the internal energy has finally relaxed
to its constant value $e_{0}$,
given by equation (\ref{stationary}).
In the minimum region $r=r_{0}$,
so the stationary value of the internal energy can be calculated directly:
  \begin{equation}
  e_{0}=\frac{c_{1}}{c_{2}+c_{3}ar_{0}{}^{2}}=\frac{1}{d_{1}}.
  \label{e0}\end{equation}
Figure \ref{fig:Internal-energy-evolution}
shows the relaxation of $e$ to the stationary value $e_{0}$.
After initial oscillations it equilibrates to a constant value
(which is then slightly disturbed by the noise influences).

Another remarkable feature of coupling the internal energy to the potential
is that equilibrium distributions can be calculated \emph{exactly}.
The standard form of stochastic differential equations
(\ref{loee}) \& (\ref{eloo})
has the following well-known stationary solution
of the corresponding Fokker Planck equation:
  \begin{equation}
  \rho(q,p)\sim e^{-\tilde{H}(q,p)},
  \end{equation}
which reduces to $\rho(q)\sim e^{-\tilde{U}(q)}$
for the distribution in configuration space,
after performing the Gaussian momentum integral.
This is an exact equilibrium solution of our nonlinear stochastic process.
Former models of active Brownian motion
had to rely on specific approximation methods,
when studying equilibrium solutions of the corresponding
Fokker Planck equation \cite{key-7}.

\section{Canonical active Brownian motion:
Coupling of $e$ to the total mechanical energy}

Combining the so far separately imposed couplings
to the kinetic, resp.\ the potential energy,
stochastic dynamics are in full generality:
  \begin{equation}
  \frac{dq_{i}}{dt}=\frac{\partial H}{\partial p_{i}}
  \label{malone one}
  \end{equation}
  \begin{equation}
  \frac{dp_{i}}{dt}
    =-f(e)\frac{\partial H}{\partial q_{i}}-g(e)
      \frac{\partial H}{\partial p_{i}}+\eta_{i}
  \end{equation}
  \begin{equation}
  \frac{de}{dt}=c_{1}-c_{2}e-c_{3}eH
  \label{malone}
  \end{equation}
with $H(q,p)=\frac{p^{2}}{2}+U(q)$ and $f(e)=1-d_{1}e$ ,
$g(e)=\gamma_{0}-d_{2}e$.
Recently another generalized model for internal energy dynamics
was introduced by Zhang et.\ al.\ \cite{Zh16}.
Whereas in their paper
the coupling of internal energy to position and velocity
is generated via an arbitrary function of $q$ and $p$,
our coupling mechanism is motivated from the idea
of exchange between internal and the full mechanical energy.
The time differential of the Hamiltonian in the deterministic case reads
  \begin{equation}
  \frac{dH}{dt}
    =-g(e)\left(\frac{\partial H}{\partial\mathbf{p}}\right)^{2}
      +d_{1}e\frac{\partial H}{\partial\mathbf{q}}
        \frac{\partial H}{\partial\mathbf{p}}
    =-\gamma_{0}p^{2}+d_{2}ep^{2}+d_{1}e\frac{dU}{dt}.
  \end{equation}
If the internal energy equilibrates after some time,
its stationary value would be
  \begin{equation}
  e(H)=\frac{c_{1}}{c_{2}+c_{3}H},
  \end{equation}
so that dynamics reduce to
  \begin{equation}
  \frac{dq_{i}}{dt}=\frac{\partial H}{\partial p_{i}}
  \label{uta}
  \end{equation}
  \begin{equation}
  \frac{dp_{i}}{dt}
  =-F(H)\frac{\partial H}{\partial q_{i}}-G(H)
  \frac{\partial H}{\partial p_{i}}+\eta_{i}
  \label{tane}
  \end{equation}
with the two dissipation functions
  \begin{equation}
  F(H)=1-\frac{d_{1}c_{1}}{c_{2}+c_{3}H}\quad,\quad G(H)
  =\gamma_{0}-\frac{d_{2}c_{1}}{c_{2}+c_{3}H}.
  \end{equation}
Stochastic equations for active Brownian motion
now truly define a \emph{canonical }system,
in the sense that all parts of the dynamics
are completely given by the Hamiltonian function,
which was not the case for separate couplings
to $\frac{p^{2}}{2}$ or $U(q)$.
Only when we consider free particles ($H=\frac{p^{2}}{2}$),
so that in equation (\ref{anders}) $\gamma(p^{2})=\gamma(H)$,
active Brownian motion in its original sense is a canonical system.
When coupling the internal energy to the Hamiltonian,
also the general case of active Brownian particles
\emph{with} interactions (i.e.\ with a potential)
represents a canonical dissipative system.
The method of coupling $e$ to the full mechanical energy
leads to a novel type of canonical dissipative system,
which has a more complex dissipative behavior
and is more general than those ones
which currently exist in the literature
\cite{mycitation,toda,qgases}.

\section{Bifurcations of Equilibria}

In this section, we investigate equilibrium solutions
of canonical active Brownian motion
and its possible bifurcation scenarios.
Bifurcation theory studies the qualitative change of solutions
for a dynamical system when parameters are varied.
\emph{Static} bifurcations are present,
e.g., when for some critical parameter values,
two equilibrium points become stable while another one loses its stability.
Apart from dealing only with stationary solutions,
\emph{dynamic} bifurcations describe
how equilibrium points can bifurcate into nonequilibrium orbits,
e.g.\ limit cycles.
We will encounter both of these bifurcation
scenarios in canonical active Brownian motion.
Special attention is given to nonequilibrium solutions,
i.e.\ situations in which $\frac{de}{dt}\neq0$.
Bifurcations of active Brownian motion
for some special case of generalized internal energy dynamics
were studied also in \cite{Zh16}.
However, no bifurcation analysis
for velocity \emph{and} space dependent internal energy equations were given,
nor was considered the case of an external harmonic potential.
We will investigate the non-equilibrium behavior of active Brownian motion,
when both kinetic \emph{and} potential energy coupling is present.

For the following explicit calculations,
we assume the harmonic external potential $U(q)=\frac{1}{2}q^{2}$.
Stochastic differential equations (\ref{malone one})-(\ref{malone}) then read
  \begin{equation}
  \frac{dq_{i}}{dt}=p_{i}\quad,\quad\frac{dp_{i}}{dt}
  =-(1-d_{1}e)q_{i}-(\gamma_{0}-d_{2}e)p_{i}+\eta_{i}
  \label{eq:one dim}
  \end{equation}
  \begin{equation}
  \frac{de}{dt}
  =c_{1}-c_{2}e-c_{3}e\frac{p^{2}}{2}-c_{4}e\frac{q^{2}}{2},
  \label{eq:one dim 2}
  \end{equation}
where we have separated the couplings of $e$
to the kinetic and the potential energy
for later analysis of special cases.

\subsection{One Dimension}

\begin{figure}
\begin{center}
\includegraphics[width=0.75\textwidth]{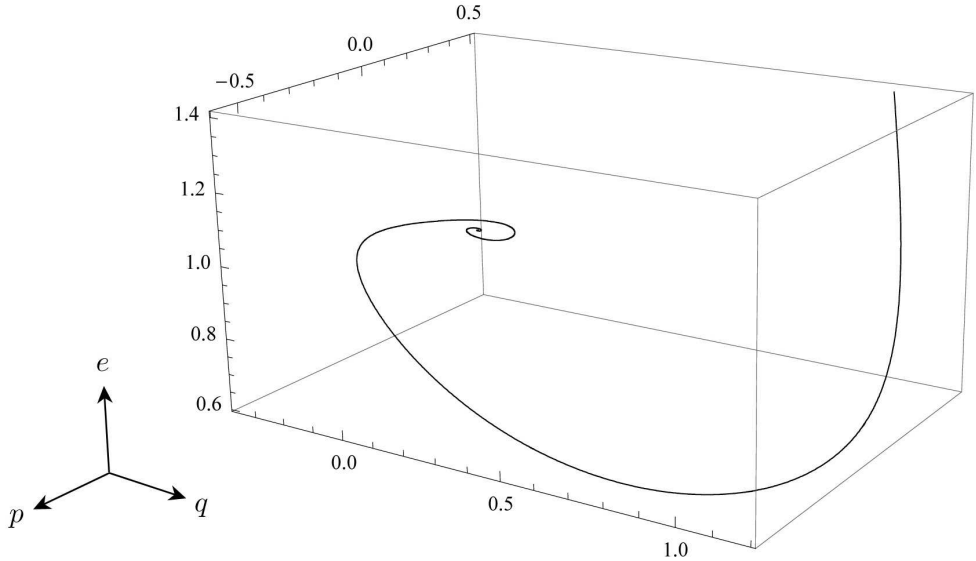}
\caption{\label{fig:traj1}
Trajectory to equilibrium (\ref{qpe:e1}),
parameters satisfying the corresponding stability conditions:
$\gamma_{0}=1$, $c_{1}=1$, $c_{2}=2$, $c_{3}=2$, $c_{4}=2$,
$d_{1}=1$, $d_{2}=1$.  (The internal energy is plotted with a factor of 2.)}
\end{center}
\end{figure}

Equilibria in one dimension for the deterministic case are
\begin{equation} \tag{e1} \label{qpe:e1}
q_{0}=0=p_{0},
\quad e_{0}=\frac{c_{1}}{c_{2}}
\end{equation}
\begin{equation} \tag{e2} \label{qpe:e2}
q_{0} = \pm \left\{ \frac{2(d_{1}c_{1}-c_{2})}{c_{4}} \right\}^{1/2},
\quad p_{0} = 0,
\quad e_{0} = \frac{1}{d_{1}}
\end{equation}
Stability conditions of these points are obtained by linearizing the system.
In equations (\ref{eq:one dim}) \& (\ref{eq:one dim 2}), we make the shift
$q\rightarrow q+q_{0}$, $p\rightarrow p+p_{0}$, $e\rightarrow e+e_{0}$,
neglecting any terms of order higher than one.
Then dynamics are given only by the Jacobian matrix,
whose eigenvalues can be studied for each of the three equilibria separately.
For equilibrium (\ref{qpe:e1}), the eigenvalues of the Jacobian are 
  \begin{equation}
  \lambda_{1}=-c_{2}\quad,\quad\lambda_{2,3}
  =-\frac{\gamma_{0}-\frac{d_{2}c_{1}}{c_{2}}}{2}
    \pm\left\{ \left(\frac{\gamma_{0}-\frac{d_{2}c_{1}}{c_{2}}}{2}\right)^{2}
    +\frac{d_{1}c_{1}}{c_{2}}-1\right\}^{\frac{1}{2}}.
  \label{eq:eigenv}
  \end{equation}
The real parts of all $\lambda_{i}$'s are negative
if $\gamma_{0}c_{2}>d_{2}c_{1}$ and $c_{2}>d_{1}c_{1}$.
Hence, equilibrium (\ref{qpe:e1}) remains stable
as long as these two conditions are fulfilled.
Figure \ref{fig:traj1} shows a trajectory in phase space $(q,p,e)$
approaching equilibrium (\ref{qpe:e1}).

The corresponding eigenvalues for equilibria (\ref{qpe:e2})
are not accessible so easily,
so we follow the Routh-Hurwitz theorem \cite{key-8},
which allows one to calculate stability regions
by analyzing only the characteristic equation of the Jacobian,
without any need for solving it.
Application of this scheme for equilibria (\ref{qpe:e2})
results in the conditions
$\gamma_{0}d_{1}>d_{2},\, c_{2}<d_{1}c_{1}$ and
$(\gamma_{0}d_{1}-d_{2})(c_{1}^{2}d_{1}^{2}+c_{1}(\gamma_{0}d_{1}-d_{2}))
>2d_{1}(c_{1}d_{1}-c_{2})$ for both of them.

\begin{figure}
\begin{center}
\includegraphics[width=0.75\textwidth]{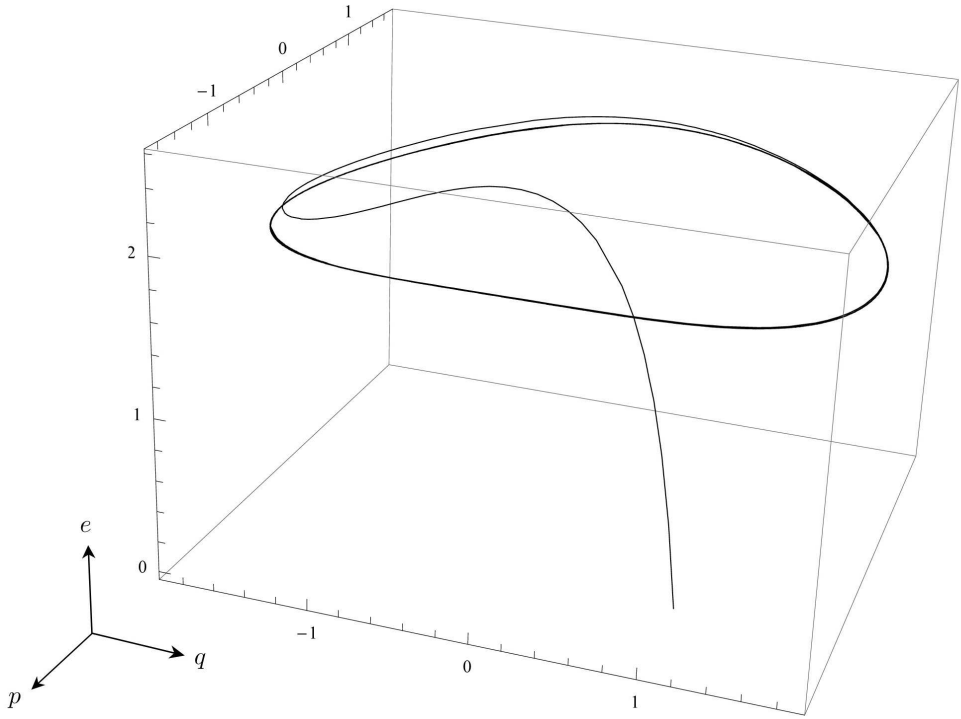}
\end{center}
\caption{\label{fig:Limit Cycle}
Limit cycle appearing after the Hopf-bifurcation point.
Parameter values are:
$\gamma_{0}=1$, $c_{1}=1$, $c_{2}=2$, $c_{3}=2=c_{4}$, $d_{1}=1$, $d_{2}=5$.
Initial conditions were chosen to be $q(0)=0$, $p(0)=-1$ and $e(0)=0$.
(The internal energy is plotted with a factor of 10.)}
\end{figure}

Two types of bifurcations can be seen, a static and a dynamic one.
First we observe a Pitchfork-bifurcation of equilibrium (\ref{qpe:e1})
into the two other ones at $c_{2}=d_{1}c_{1}$:
As long as $c_{2}>d_{1}c_{1}$ is fulfilled,
equilibrium (\ref{qpe:e1}) is stable,
while equilibria (\ref{qpe:e2}) do not exist.
For parameter regions where $c_{2}<d_{1}c_{1}$,
equilibria (\ref{qpe:e2}) are stable,
while equilibrium (\ref{qpe:e1}) becomes unstable.
Apart from this static bifurcation we observe a dynamic bifurcation,
namely a Hopf-bifurcation, at $\gamma_{0}c_{2}=d_{2}c_{1}$:
Suppose we fix the condition $c_{2}>d_{1}c_{1}$,
by looking at the eigenvalues (\ref{eq:eigenv}),
we see the appearance of a purely imaginary pair
of eigenvalues $\lambda_{2,3}=\pm i\omega$
at the critical point $\gamma_{0}c_{2}=d_{2}c_{1}$
(while $\mathrm{Re}\lambda_{1}\neq0$).
This is the general condition for the existence of a Hopf-bifurcation,
which describes equilibria bifurcating into periodic solutions \cite{key-9}.
Equilibrium (\ref{qpe:e1}) bifurcates into limit cycles at the Hopf bifurcation
point $\gamma_{0}c_{2}=d_{2}c_{1}$.
One example of these limit cycles is shown in figure \ref{fig:Limit Cycle}.

\subsection{Arbitrary Dimensions}

A similar study can be made for arbitrary dimensions $n$.
First we reduce the $(2n+1)$-dimensional system
(\ref{eq:one dim}) \& (\ref{eq:one dim 2})
to four dimensions by transforming to the variables
$K=\frac{p^{2}}{2}$, $U=\frac{q^{2}}{2}$ and $S=\mathbf{qp}$.
Evolution equations in its deterministic form therefore transform to
  \begin{equation}
  \frac{dK}{dt}=-2(\gamma_{0}-d_{2}e)K-(1-d_{1}e)S
  \quad,\quad\frac{dU}{dt}=S
  \end{equation}
  \begin{equation}
  \frac{dS}{dt}=2K-(\gamma_{0}-d_{2}e)S-2(1-d_{1}e)U,
  \end{equation}
  \begin{equation}
  \frac{de}{dt}=c_{1}-c_{2}e-c_{3}eK-c_{4}eU. 
  \end{equation}
The following are the three possible equilibrium points:
  \begin{equation} \tag{E1} \label{svke:e1}
  S_{0}=K_{0}=U_{0}=0, \quad e_{0}=\frac{c_{1}}{c_{2}}
  \end{equation}
  \begin{equation} \tag{E2} \label{svke:e2}
  S_{0}=K_{0}=0, \quad U_{0}=\frac{c_{1}d_{1}-c_{2}}{c_{4}},
                 \quad e_{0}=\frac{1}{d_{1}}
  \end{equation}
  \begin{equation} \tag{E3} \label{svke:e3}
  S_{0} = 0, \quad
  K_{0} = \left(1-\frac{d_{1}\gamma_{0}}{d_{2}}\right)U_{0}, \quad
  U_{0} = \frac{d_{2}c_{1}-c_{2}\gamma_{0}}{
        \gamma_{0}c_{3}\left(1-\frac{d_{1}\gamma_{0}}{d_{2}}\right)
              +\gamma_{0}c_{4}},\quad
  e_{0} = \frac{\gamma_{0}}{d_{2}}.
  \end{equation}
Before addressing bifurcation theory,
we shortly comment on how these equilibrium points
of canonical active Brownian motion contain equilibria
of the separately coupled versions.
The equilibrium solution of original active Brownian motion
(coupling of $e$ only to the kinetic energy)
for a harmonic potential \cite{key-2}
is identical to the equilibrium point (\ref{svke:e3}),
if $d_{1}=0$ and $c_{4}=0$.
The case of internal energy coupling only to the potential,
studied in Section \ref{sec:potential coup},
is identical to equilibrium (\ref{svke:e3}),
also for nonvanishing $d_{2}$ and $c_{3}$.
Active Brownian motion in its fully coupled form
therefore shares the same equilibrium solution with the system
when coupled only to the potential.
Therefore the same trajectories
as already shown in figure \ref{fig:diffusion circle} can be seen.

\begin{figure}
\begin{center}
\includegraphics{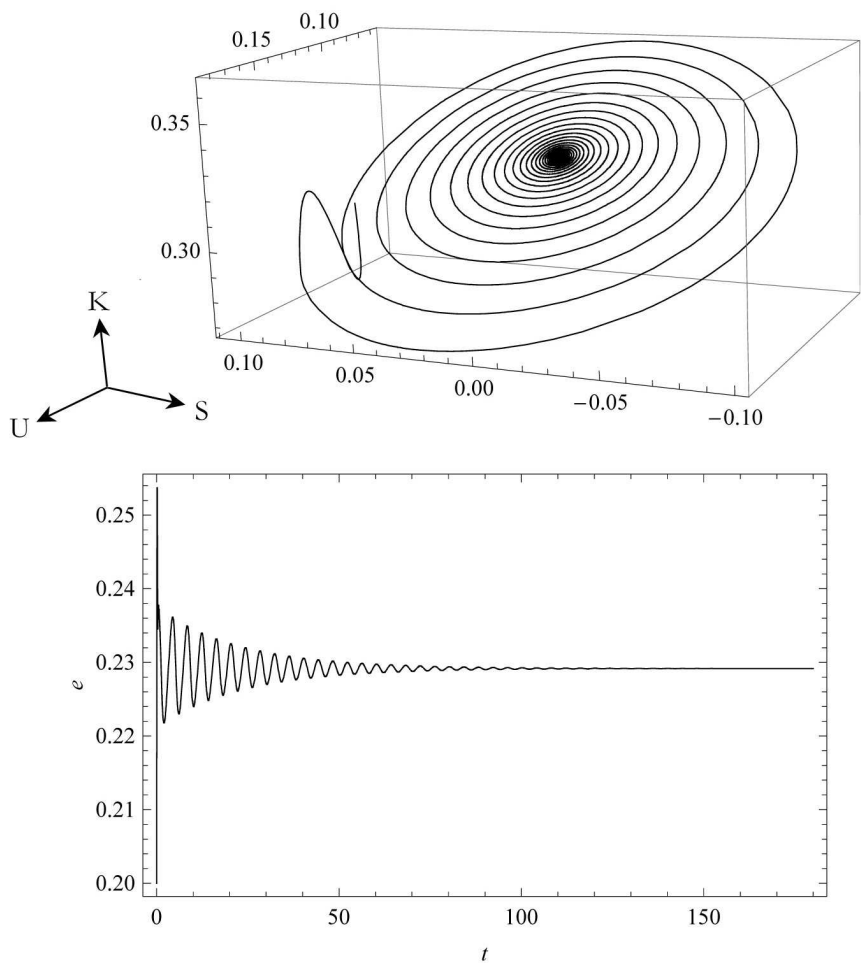}
\end{center}
\caption{\label{fig:traj3}
Trajectory in 4-dimensional phase space to equilibrium (\ref{svke:e3})
for the following choice of parameters:
$\gamma_{0}=11$, $c_{1}=4$, $c_{2}=0.01$,
$c_{3}=129$, $c_{4}=31$, $d_{1}=0.01$, $d_{2}=48$
and initial conditions $S(0)=0.1$, $U(0)=0.1$, $K(0)=0.1$, $e(0)=0.2$.
(The kinetic energy $K$ is plotted with a factor of 3.)}
\end{figure}

We now repeat our linearization procedure concerning stability
properties and bifurcations. Eigenvalues of the Jacobian associated
with equilibrium (\ref{svke:e1}) can be calculated straightforwardly:
  \begin{equation}
  \lambda_{1}=-c_{2}\quad,\quad\lambda_{2}
  =\frac{c_{1}d_{2}-c_{2}\gamma_{0}}{c_{2}}
  \end{equation}
  \begin{equation}
  \lambda_{3,4}
  =\frac{c_{1}d_{2}-c_{2}\gamma_{0}}{c_{2}}
  \pm\left\{ \left(\gamma_{0}-\frac{d_{2}c_{1}}{c_{2}}\right)^{2}
  +4\left(\frac{d_{1}c_{1}}{c_{2}}-1\right)\right\}^{\frac{1}{2}}.
  \label{eq:eigenv2}
  \end{equation}
The real parts are negative if the conditions
$\gamma_{0}c_{2}>d_{2}c_{1}$ and $c_{2}>d_{1}c_{1}$ are fulfilled
(these are the same conditions
as for equilibrium (\ref{qpe:e1}) in one dimension).
The stability analysis for equilibrium (\ref{svke:e2})
can be carried out via the Routh-Hurwitz theorem
and it leads to the same conditions
as for equilibrium (\ref{qpe:e2}) in one dimension.
Equilibrium (\ref{svke:e3}) is hard to analyze
even with the Routh-Hurwitz procedure,
so that we have no information about stability regions
for this third stationary solution at all.
Nonetheless we are able to find parameter values by hand
which lead to trajectories approaching it slowly,
as is shown in Figure \ref{fig:traj3}.
Concerning the bifurcation behavior of our system,
we observe two collisions of equilibria
at critical parameter values of $d_{2}$
(which we choose as our bifurcation parameter in the following).

\begin{figure}
\begin{center}
\includegraphics[width=\textwidth]{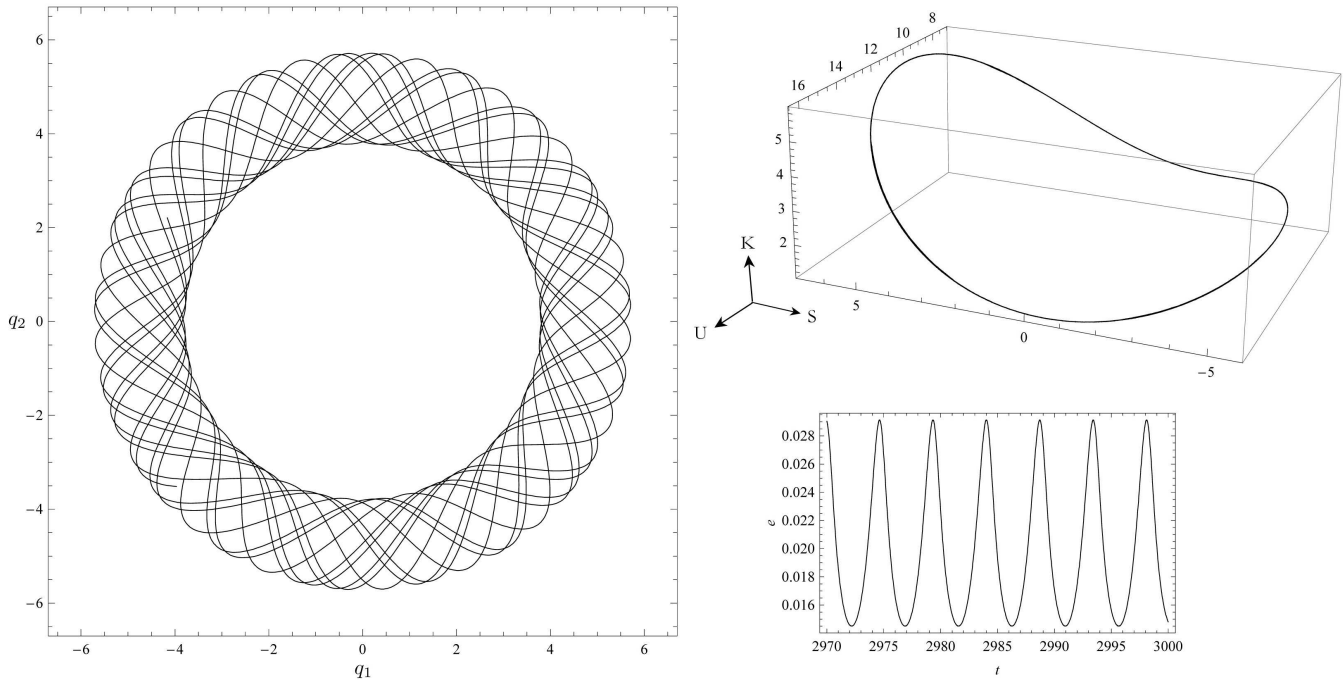}
\end{center}
\caption{\label{fig:rosette}
Limit cycle in the phase space of variables $(K,U,S,e)$
and its corresponding quasi-periodic motion in configuration space
for $n=2$ after the transient phase.
Parameter values are $\gamma_{0}=1$, $c_{1}=1$, $c_{2}=2$, $c_{3}=4$,
$c_{4}=4$, $d_{1}=49$, $d_{2}=50$
and initial conditions $S(0)=0$, $U(0)=0$, $K(0)=1$, $e(0)=0$.
Corresponding initial conditions in $(q_{i},p_{i},e)$:
$q_{1}(0)=0=q_{2}(0)$, $p_{1}(0)=-1=p_{2}(0)$, $e(0)=0$.
(The kinetic energy $K$ is plotted with a factor of 3.)}
\end{figure}

Such collisions are usually associated with Fold-bifurcations.
Its general property is the existence
of a simple zero eigenvalue of the Jacobian.
This happens for equilibrium (\ref{svke:e1})
when $d_{2}=\frac{c_{2}\gamma_{0}}{c_{1}}$,
so that $\lambda_{2}$ vanishes.
At the Fold-bifurcation point $d_{2}=\frac{c_{2}\gamma_{0}}{c_{1}}$,
equilibrium (\ref{svke:e1}) becomes identical with equilibrium (\ref{svke:e3}).
Before the bifurcation, when $d_{2}<\frac{c_{2}\gamma_{0}}{c_{1}}$,
equilibrium (\ref{svke:e1}) is stable, while (\ref{svke:e3}) does not exist,
since the potential $U=\frac{1}{2}q^{2}$ has to be positive.
After the Fold-point, when $d_{2}>\frac{c_{2}\gamma_{0}}{c_{1}}$,
equilibrium (\ref{svke:e1}) is unstable,
and (\ref{svke:e3}) is observed to be stable
(see Figure \ref{fig:traj3} for one example).

Another Fold-bifurcation is present at $d_{2}=\gamma_{0}d_{1}$,
which is a bifurcation of equilibrium (\ref{svke:e2}).
Although three of its eigenvalues are not accessible in a treatable way,
one of them is in a simple form,
namely $\lambda=\frac{2(d_{2}-d_{1}\gamma_{0})}{d_{1}}$.
It vanishes for $d_{2}=\gamma_{0}d_{1}$ and we see the collision
of equilibrium (\ref{svke:e2}) with (\ref{svke:e3}) at this point.
Before this critical value, when $d_{2}<\gamma_{0}d_{1}$,
equilibrium (\ref{svke:e2}) is stable, while (\ref{svke:e3}) does not exist,
since the kinetic energy $K=\frac{1}{2}p^{2}$ has to be positive.
After the bifurcation point, when $d_{2}>\gamma_{0}d_{1}$,
equilibrium (\ref{svke:e2}) is unstable,
while (\ref{svke:e3}) can be observed to be stable
(see again Figure \ref{fig:traj3}).

Apart from these two collisions of equilibria,
a Fold-Hopf bifurcation is present.
Remember that in one dimension we observed a Hopf bifurcation
at some critical point in parameter space.
In arbitrary dimensions,
if the condition $c_{2}>d_{1}c_{1}$ is fulfilled,
the eigenvalues $\lambda_{3,4}$ of equilibrium (\ref{svke:e1})
become a purely conjugate complex pair $\lambda_{3,4}=\pm i\omega$
at $d_{2}=\frac{\gamma_{0}c_{2}}{c_{1!}}$
and $\lambda_{2}$ vanishes exactly.
This zero-pair constellation at some critical point in parameter space
is the general condition for a Fold-Hopf-bifurcation.
To this type of bifurcation are associated various non-equilibrium dynamics,
ranging from motions on tori to heteroclinic orbits \cite{key-9}.
We do not want to address the analysis of this Fold-Hopf-bifurcation here,
instead focusing on the interesting case,
when stable limit cycles are present in our 4-dimensional system.
Figure \ref{fig:rosette} shows a limit cycle in phase space $(K,U,S,e)$.
This periodic motion corresponds to a quasi-periodic motion
in configuration space, as is shown for $n=2$.
Adding noise to our system doesn't change the qualitative behavior,
only the direction of movement can change at some instances.

Concluding this analysis of non-equilibrium behavior,
we want to emphasize the various possibilities
that arise from studying parameter regions where $e$ is not constant,
but heavily oscillating.
Far away from stationary solutions,
our system shows to have quasi-periodic dynamics.

\section{Synopsis and Outlook}

We have formulated a generalized version of active Brownian motion
in the sense that we allow not only couplings
of the internal energy to the kinetic energy,
but also to the potential energy and more generally to the Hamiltonian.
The latter case gives rise to a canonical dissipative system.
Analysis of stationary points and its possible bifurcations
into non-equilibrium solutions
reveals a rich dynamical structure in parameter regions
far away from equilibrium.
Explicit numerical studies of all bifurcation scenarios
will be left to future computer art;
nevertheless we were able to find interesting quasi-periodic dynamics by hand.
The various non-equilibrium phenomena
may be relevant in future applications to real systems,
which do not rely on the assumption of stationary internal energies.
Moreover, many particle systems with mutual interactions
can be studied within the scheme of canonical active Brownian motion.
The self-organizing properties e.g.\ of swarms
will be of special interest for upcoming papers,
when applying our model for a system composed of many particles.

\begin{acknowledgments}
We thank Harald Grosse, Markus Heinzle and Josef Hofbauer
for advice and valuable discussions.
We are grateful for financial support
within the Agreement on Cooperation
between the Universities of Vienna and Zagreb.
\end{acknowledgments}

\end{document}